\documentclass[twocolumn,prb,superscriptaddress,preprintnumbers,amsmath,amssymb,showpacs,floatfix]{revtex4}

\usepackage{graphicx}
\usepackage{dcolumn}
\usepackage{bm}
\usepackage{color}

\begin{document}

\title{Epitaxy, stoichiometry, and magnetic properties of Gd--doped EuO films on YSZ (001)}

\author{R. Sutarto}
 \affiliation{II. Physikalisches Institut, Universit\"{a}t zu K\"{o}ln,
 Z\"{u}lpicher Str. 77, 50937 K\"{o}ln, Germany}
\author{S. G. Altendorf}
 \affiliation{II. Physikalisches Institut, Universit\"{a}t zu K\"{o}ln,
 Z\"{u}lpicher Str. 77, 50937 K\"{o}ln, Germany}
\author{B. Coloru}
 \affiliation{II. Physikalisches Institut, Universit\"{a}t zu K\"{o}ln,
 Z\"{u}lpicher Str. 77, 50937 K\"{o}ln, Germany}
\author{M. Moretti Sala}
 \affiliation{II. Physikalisches Institut, Universit\"{a}t zu K\"{o}ln,
 Z\"{u}lpicher Str. 77, 50937 K\"{o}ln, Germany}
\author{T. Haupricht}
 \affiliation{II. Physikalisches Institut, Universit\"{a}t zu K\"{o}ln,
 Z\"{u}lpicher Str. 77, 50937 K\"{o}ln, Germany}
\author{C.~F.~Chang}
 \affiliation{II. Physikalisches Institut, Universit\"{a}t zu K\"{o}ln,
 Z\"{u}lpicher Str. 77, 50937 K\"{o}ln, Germany}
\author{Z.~Hu}
 \affiliation{II. Physikalisches Institut, Universit\"{a}t zu K\"{o}ln,
 Z\"{u}lpicher Str. 77, 50937 K\"{o}ln, Germany}
\author{C. Sch{\"u}{\ss}ler-Langeheine}
 \affiliation{II. Physikalisches Institut, Universit\"{a}t zu K\"{o}ln,
 Z\"{u}lpicher Str. 77, 50937 K\"{o}ln, Germany}
\author{N. Hollmann}
 \affiliation{II. Physikalisches Institut, Universit\"{a}t zu K\"{o}ln,
 Z\"{u}lpicher Str. 77, 50937 K\"{o}ln, Germany}
\author{H. Kierspel}
 \affiliation{II. Physikalisches Institut, Universit\"{a}t zu K\"{o}ln,
 Z\"{u}lpicher Str. 77, 50937 K\"{o}ln, Germany}
 \author{J. A. Mydosh}
 \affiliation{II. Physikalisches Institut, Universit\"{a}t zu K\"{o}ln,
 Z\"{u}lpicher Str. 77, 50937 K\"{o}ln, Germany}
\author{H. H. Hsieh}
 \affiliation{Chung Cheng Institute of Technology,
 National Defense University, Taoyuan 335, Taiwan}
\author{H.-J. Lin}
 \affiliation{National Synchrotron Radiation Research Center,
 101 Hsin-Ann Road, Hsinchu 30077, Taiwan}
\author{C. T. Chen}
 \affiliation{National Synchrotron Radiation Research Center,
 101 Hsin-Ann Road, Hsinchu 30077, Taiwan}
\author{L. H. Tjeng}
 \affiliation{II. Physikalisches Institut, Universit\"{a}t zu K\"{o}ln,
 Z\"{u}lpicher Str. 77, 50937 K\"{o}ln, Germany}

\date{\today}

\begin{abstract}
We have succeeded in preparing high-quality Gd--doped
single--crystalline EuO films. Using Eu--distillation--assisted
molecular beam epitaxy and a systematic variation in the Gd and
oxygen deposition rates, we have been able to observe sustained
layer--by--layer epitaxial growth on yttria--stabilized cubic
zirconia (001). The presence of Gd helps to stabilize the
layer--by--layer growth mode. We used soft x--ray absorption
spectroscopy at the Eu and Gd $M_{4,5}$ edges to confirm the
absence of Eu$^{3+}$ contaminants and to determine the actual Gd
concentration. The distillation process ensures the absence of
oxygen vacancies in the films. From magnetization measurements we
found the Curie temperature to increase smoothly as a function of
doping from 70 K up to a maximum of 125 K. A threshold behavior
was not observed for concentrations as low as 0.2\%.
\end{abstract}

\pacs{68.55.--a, 75.70.Ak, 78.70.Dm, 79.60.Dp, 81.15.Hi}

\maketitle

\section{Introduction}
Renewed research interest in europium oxide (EuO) thin films has
been arising in recent years with the goal to utilize its
extraordinary properties for spintronic
applications.\cite{schmehl07a,panguluri08a,santos08a} At
room--temperature stoichiometric bulk EuO is a paramagnetic
semiconductor with a band gap of about 1.2~eV and upon cooling, it
orders ferromagnetically with a Curie temperature ($T_C$) of
69~K.\cite{mauger86a,tsuda91a} In the ferromagnetic state, the
spin--up and spin--down conduction bands experience a rather large
splitting of about 0.6~eV due to the indirect exchange interaction
between the magnetic moment of the localized Eu$^{2+}$ $4f^{7}$
electrons and the delocalized $5d$--$6s$ ones.\cite{steeneken02a}
It has been inferred that, owing to this spin--splitting of the
conduction band, extra electrons introduced in EuO will be
practically 100\% spin polarized,\cite{steeneken02a} and it was
recently demonstrated that this is indeed the case for
lanthanum--doped EuO$_{1-x}$.\cite{schmehl07a}

Another eyes catching property of EuO is that the $T_C$ can be
significantly enhanced by electron doping, for instance, by
substituting the Eu with trivalent rare
earths.\cite{mauger86a,tsuda91a} Numerous studies have been
carried out to characterize and optimize the doping--induced
magnetic properties. The results, however, vary appreciably. For
Gd--doped EuO, for example, the reported optimum $T_C$ value
ranges from 115 to 148
K,\cite{shafer68a,ahn68a,samokhvalov72a,samokhvalov74a,schoenes74a,mauger78a,meier79a,godart80a,mauger80a,rho02a,matsumoto04a}
and even up to 170 K.\cite{ott06a} The temperature dependence of
the magnetization also differs from report to report, most of them
showing very little resemblance to a Brillouin function as
expected for a Heisenberg system. It is also not clear whether a
threshold of the Gd concentration (if any) exists for $T_C$ starts
to
increase.\cite{samokhvalov72a,samokhvalov74a,schoenes74a,meier79a,mauger80a,rho02a,matsumoto04a}
It was asserted that much of these uncertainties are probably
caused by problems with stoichiometry.\cite{tsuda91a,schoenes74a}
Indeed, oxygen deficiencies in Gd--free EuO samples already lead
to enhancements of $T_C$ up to 140--150
K.\cite{massenet74a,samokhvalov78a,matsumoto04a} In fact, it was
also suggested that even the actual Gd concentration was not known
accurately.\cite{tsuda91a}

We have recently developed the so--called
Eu--distillation--assisted molecular beam epitaxy (MBE) procedure
to prepare single--crystalline and highly stoichiometric EuO thin
films on yttria--stabilized cubic zirconia (YSZ)
substrates.\cite{sutarto09a} Our objective now is to use this
procedure as a starting point for obtaining high quality Gd--doped
EuO samples. In this study we will investigate in detail whether
or not the Eu--distillation process indeed allows for the growth
of Gd--doped films free from Eu$^{3+}$ contaminants, oxygen
vacancies, Eu metal clusters, and Gd$_2$O$_3$ phases as well.
Concerning the growth process itself, we would like to know
whether the layer--by--layer growth mode observed for pure EuO on
YSZ can also be maintained in the presence of Gd co--deposition.
We will use soft x--ray absorption spectroscopy at the Eu and Gd
$M_{4,5}$ edges to obtain a reliable determination of the actual
Gd concentration. Our goal is then to establish the magnetic
properties of the Gd--doped EuO system using well--defined films.

\section{Experiment}
The Gd--doped EuO films were grown in an ultra--high--vacuum (UHV)
MBE facility with a base pressure of 2$\times$10$^{-10}$ mbar.
Epi--polished YSZ single crystals from SurfaceNet GmbH were used
as substrates. The substrate crystal structure is
calcium--fluorite type and the surface normal of the substrate is
the [001] direction. The lattice constant of YSZ is 5.142
\AA,\cite{ingel86a,yashima94a} very close to the 5.144 \AA\ value
for EuO at room temperature.\cite{henrich94a} Prior to growth, the
YSZ substrates were annealed \textit{in situ} at $T=600^{\circ}$C
in an oxygen atmosphere of $5\times10^{-7}$ mbar for at least 120
min in order to obtain clean and well-ordered substrate surfaces.

\begin{figure*}[t]
\includegraphics*[width = 17 cm] {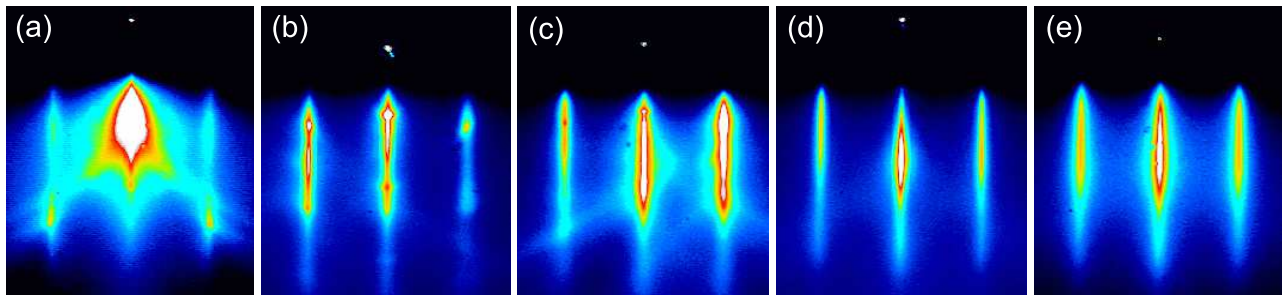}
\caption{\label{Fig_1} (Color online) RHEED photographs of (a)
clean and annealed YSZ (001), and Gd--doped EuO films with Gd
concentrations of (b) 0.2\%, (c) 2.0\%, (d) 7.7\%, and (e) 20\%.
The films were grown for 100 min using a 8.1--8.2~\AA/min Eu flux
rate and a $8\times10^{-8}$ mbar oxygen pressure. The Gd flux
rates were 0.006, 0.056, 0.51, and 0.7~\AA/min for (b)--(e),
respectively. The RHEED electron energy was 20 keV with the beam
incident along the [100] direction.}
\end{figure*}

The films were prepared by simultaneously depositing Eu and Gd
metals on top of YSZ substrates under oxygen atmosphere. High
purity Eu metal from AMES Laboratory was sublimated from an EPI
effusion cell with a BN crucible at temperatures between 525 and
545$^{\circ}$C. Gd metal from Smart--Elements Co. was evaporated
from a commercial Luxel Corporation RADAK--I Knudsen cell with a
molybdenum--insert containing Al$_2$O$_3$ crucible at temperatures
between 1100 and 1350$^{\circ}$C. Proper degassing of the Eu and
Gd materials ensured that during the film deposition the
background pressure was kept below 5$\times$10$^{-9}$ mbar. The
Eu--deposition rate of 8.1--8.2~\AA/min and the Gd rate of
0.006--0.8~\AA/min were calibrated using a quartz--crystal
monitor, which was moved to the sample--growth position prior and
after each growth. Molecular oxygen was supplied through a leak
valve, and its pressure (4--10$\times$10$^{-8}$ mbar) was
monitored using an ion--gauge and a mass--spectrometer. The
substrates were kept at $T=400^{\circ}$C during growth and all
films were grown for 100 min.

The MBE facility is equipped with the EK--35--R reflection
high-energy electron diffraction (RHEED) system from STAIB
Instruments for \textit{online} monitoring of the \textit{in situ}
growth. The facility is connected to an UHV $\mu$--metal
photoemission chamber supplied with a Vacuum Generators Scientific
T191 rear--view low--energy electron diffraction (LEED) system for
further \textit{in situ} structural characterization. The facility
is also attached to a separate UHV chamber for the evaporation of
aluminum as protective capping layer of the air--sensitive
Gd--doped EuO films. It allows us to perform \textit{ex situ}
characterizations using x--ray reflectivity (XRR), superconducting
quantum interference device (SQUID), and x--ray absorption
spectroscopy (XAS). The XRR measurements were carried out using a
Siemens D5000 diffractometer. The thicknesses of the films are
about 200, 425, and 515~\AA~for the oxygen pressure of 4, 8, and
10$\times$10$^{-8}$ mbar, respectively. The thickness of the
aluminum capping is about 20--40~\AA. The magnetic properties of
the films were determined using a Quantum Design MPMS--XL7 SQUID
magnetometer. The XAS measurements were performed at the Dragon
beamline of the National Synchrotron Radiation Research Center
(NSRRC) in Taiwan. The spectra were recorded using the total
electron yield method and the photon--energy resolution at the Eu
and Gd $M_{4,5}$ edges ($h\nu \approx 1100$--1235~eV) was set at
$\approx$ 0.6~eV.

\begin{figure*}[t]
\includegraphics*[width = 17 cm] {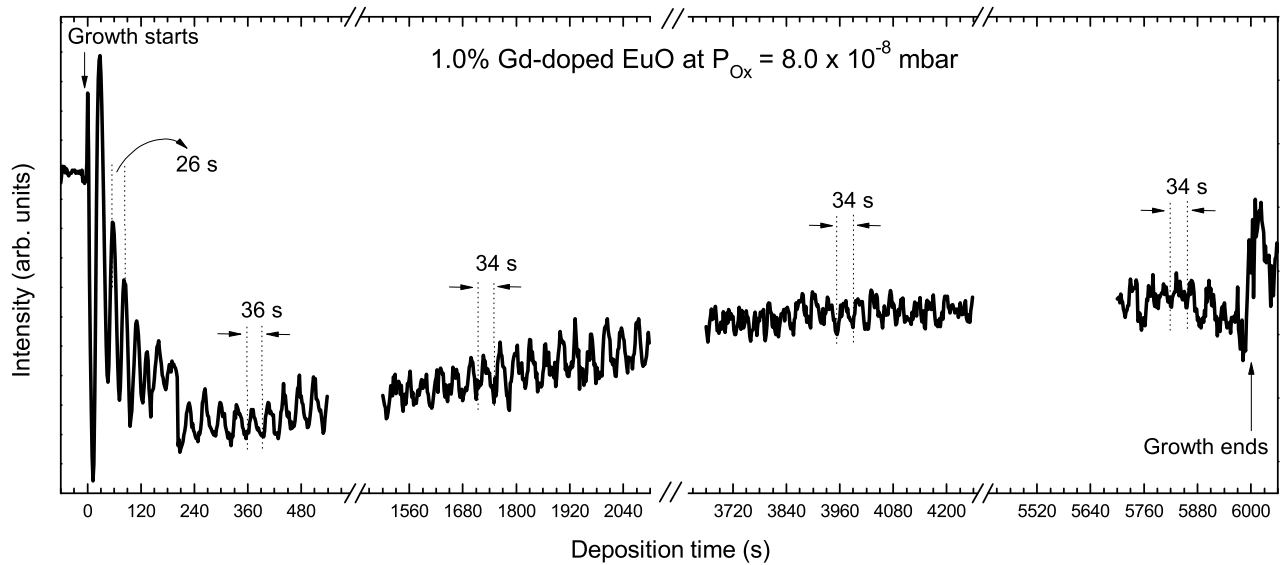}
\caption{\label{Fig_2} RHEED intensity oscillations of the
specularly reflected electron beam, detected during deposition of
a 1\% Gd--doped EuO film on YSZ (001) grown using a
8.1--8.2~\AA/min Eu flux rate and a $8\times10^{-8}$ mbar oxygen
pressure $P_{\rm Ox}$.}
\end{figure*}

\begin{figure}[t]
\includegraphics*[scale = 0.3] {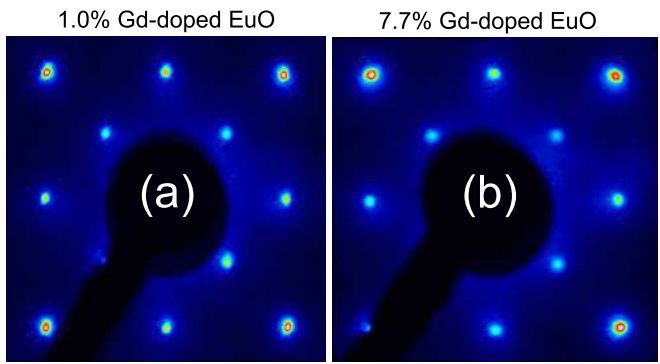}
\caption{\label{Fig_3} (Color online) LEED photographs of
epitaxial Gd--doped EuO films on YSZ (001), grown for 100 min
using a 8.1--8.2~\AA/min Eu flux rate and a $8\times10^{-8}$ mbar
oxygen pressure. The Gd concentration is (a) 1.0\% and (b) 7.7\%.
The patterns were recorded at an electron beam energy of
approximately 216 eV.}
\end{figure}

\section{Results and Discussion}

\subsection{Growth of Gd--doped EuO films}
\textit{In situ} RHEED was employed to monitor the growth quality
of Gd--doped EuO thin films. The RHEED photograph of the clean and
annealed YSZ (001) before growth is displayed in
Fig.~\ref{Fig_1}(a). Figures~\ref{Fig_1}(b)--\ref{Fig_1}(e) show
the photographs of Gd--doped EuO films grown with Gd
concentrations of (b) 0.2\%, (c) 2.0\%, (d) 7.7\%, and (e) 20\%.
The films were prepared at 400$^{\circ}$C under an oxygen pressure
of $8\times10^{-8}$ mbar and a Eu flux rate of 8.1--8.2~\AA/min.
The Gd flux rates were 0.006, 0.056, 0.51, and 0.7~\AA/min for
(b)--(e), respectively. We will describe later how the actual Gd
concentration was determined using XAS at the Eu and Gd $M_{4,5}$
edges.

The important result is that epitaxial growth of Gd--doped EuO
films has been achieved for a wide range of Gd concentrations,
even up to 20\% doping. The distance between the streaks of the
Gd--doped EuO films is always very similar to that of the YSZ
substrate, confirming that the in-plane lattice parameters of
Gd--doped EuO and YSZ are closely matched. The sharp streaks in
the RHEED patterns after 100 min of growth can be taken as an
indication for the smoothness of the film surface.

We were also looking for RHEED intensity oscillations during the
growth of the Gd--doped EuO films. Such oscillations then indicate
the occurrence of a layer--by--layer growth mode, which is
important to obtain high--quality smooth films. In our previous
study on undoped EuO films grown on YSZ (001), we always observed
five to six oscillations after the start of the
growth.\cite{sutarto09a} The oscillation period is determined by
the Eu flux rate and is independent of the oxygen pressure
(provided that it does not exceed a certain critical value above
which Eu$^{3+}$ could be formed). We found that this is a unique
feature for the growth on YSZ, and we were able to attribute this
phenomenon to the fact that the YSZ substrate acts as a source for
oxygen, which very remarkably, is capable in oxidizing Eu to
Eu$^{2+}$ but not to Eu$^{3+}$.\cite{sutarto09a} The period was
about 25 s for an Eu flux rate of
8.1--8.2~\AA/min.\cite{sutarto09a} Also here for the Gd--doped
films we found five to six oscillations during the initial stages
of growth. The period is approximately 26 s when the Gd
concentration is very low, and decreases to 22 s for films
containing 20\% Gd, suggesting that the period is roughly
inversely proportional to the sum of the Eu and Gd flux rates.
These results provide a consistent picture in that the initial
stages of growth are governed by the oxygen supply from the YSZ
substrate.

With regard to sustained growth, we found in our previous
study\cite{sutarto09a} that long lasting RHEED intensity
oscillations can be observed for undoped EuO films on YSZ (001),
provided that the oxygen pressure in the MBE chamber is close but
not exceeding the critical value above which Eu$^{3+}$ ions are
formed. The oscillation time is then no longer determined by the
Eu flux rate, but by the oxygen pressure. The growth process,
therefore, involves the re--evaporation of the excess Eu into the
vacuum. Also here for the Gd--doped films we make use of this
so--called Eu--distillation--assisted growth process. Again, using
oxygen pressures slightly lower than the critical value, we are
able to observe prolonged oscillations. Figure~\ref{Fig_2} shows
the time dependence of the RHEED intensity of the specularly
reflected beam during the deposition of a 1.0\% Gd--doped EuO
film. The initial five to six oscillations with the periods of 26
s are followed by numerous oscillations with a period of about
34.5 s for the entire duration of growth. This demonstrates that a
Gd--doped EuO film can be prepared epitaxially in a
layer--by--layer fashion.

We find those prolonged RHEED intensity oscillations to occur not
only for the lowest Gd concentrations but also for concentrations
as high as 8\%. Although we have not carried out a systematic
study, we have indications that the range of oxygen pressures for
which the oscillations can be observed is larger for the Gd--doped
EuO films than for the pure EuO. So it seems that the presence of
Gd does help to stabilize the two--dimensional layer--by--layer
growth mode. It is tempting to speculate that perhaps the Gd ions
could act as nonmobile nucleation sites since the Gd vapor
pressure is extremely low. Those nucleation sites then would
increase the step density and make it oscillate for every
formation of a new layer.

To check the surface structure, LEED experiments were performed
after the growth of Gd--doped EuO films was completed.
Figure~\ref{Fig_3} depicts examples of the LEED photographs of the
films with Gd concentration of 1.0\% for (a) and 7.7\% for (b).
The patterns were recorded at an electron beam energy of
approximately 216 eV. Since these Gd--doped films are not as
insulating as the pure EuO film,\cite{sutarto09a} we were also
able to record the LEED patterns at lower energies, down to about
98 eV (not shown). Figure~\ref{Fig_3} reveals a perfect (001)
surface of the rock-salt structure, fully consistent with the
RHEED results.

\subsection{Stoichiometry and Gd concentration}

\begin{figure}[t]
\includegraphics*[scale = 0.35] {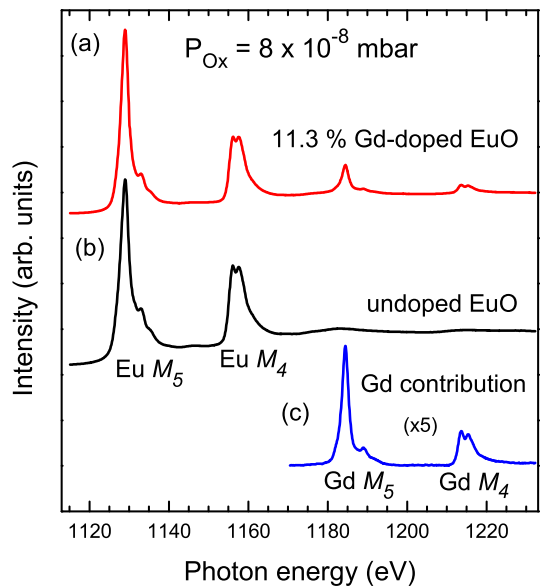}
\caption{\label{Fig_4} (Color online) Eu and Gd $M_{4,5}$ ($3d
\rightarrow 4f$) XAS spectra of (a) a 11.3\% Gd--doped EuO film
and (b) an undoped EuO film at 300~K. The net Gd $M_{4,5}$
contribution (c) is obtained after subtracting the EXAFS of the
undoped EuO from the 11.3\% Gd--doped EuO spectrum.}
\end{figure}

We have investigated the stoichiometry and Gd concentration of the
Gd--doped EuO films using \textit{ex situ} soft XAS measurements.
XAS is an element--specific method that is very sensitive to
chemical state of a probed ion.\cite{thole85a,goedkoop88a} Since
Gd--doped EuO films are highly susceptible towards further
oxidation in air, we have capped the films with a protective
aluminum layer of 20--40~\AA~before taking them out of the MBE
system and subsequent transport to the synchrotron
facility.\cite{sutarto09a} Figures~\ref{Fig_4}(a)
and~\ref{Fig_4}(b) depict the Eu and Gd $M_{4,5}$ ($3d
\rightarrow4f$) XAS spectra of a 11.3\% Gd--doped EuO film and an
undoped EuO film, respectively. Both films show identical line
shapes in their Eu $M_{4,5}$ spectra. Comparing to the theoretical
spectra of Eu$^{2+}$ and Eu$^{3+}$,\cite{thole85a,goedkoop88a} we
can directly conclude that the experimental spectra represent
exclusively Eu$^{2+}$ ions. There are no extra peaks or shoulders
which otherwise could indicate the presence of Eu$^{3+}$
species.\cite{panguluri08a,santos08a} We have also carried out XAS
measurements on films with other Gd concentrations, and can
confirm that our films are completely free from Eu$^{3+}$
contaminants.

The Gd $M_{4,5}$ spectrum has all the characteristics of a
$3d^{10}4f^{7}$$\rightarrow$$3d^{9}4f^{8}$
transition,\cite{thole85a,goedkoop88a,rudolf92a} similar to that
of the Eu$^{2+}$ $M_{4,5}$ one. Thus, having an identical 4$f$
configuration and very similar spectral line shapes as well as
photoabsorption cross--sections, we can use the XAS as a simple
and reliable method to deduce the Gd concentration in the
films.\cite{ott06a} Figure~\ref{Fig_4}(c) shows the net
contribution of the Gd spectrum after subtracting the extended
x--ray absorption fine structure (EXAFS) of the undoped EuO
[Fig.~\ref{Fig_4}(b)] from the Gd--doped EuO [Fig.~\ref{Fig_4}(a)]
spectrum in the Gd $M_{4,5}$ energy range. The Gd spectrum is
displayed with a magnification of a factor of 5 for clarity. Now
one can directly see that the Eu and Gd spectral line shapes are
indeed identical. Furthermore, a Gd/Eu ratio is obtained by
dividing the integrated Gd $M_5$ and Eu $M_5$ intensities. In the
example of Fig.~\ref{Fig_4}, a Gd/Eu ratio of 12.7\% is extracted,
corresponding to a Gd concentration of $x$ = 11.3\% in the
Eu$_{1-x}$Gd$_x$O chemical formula.

\begin{figure}[t]
\includegraphics*[scale = 0.35] {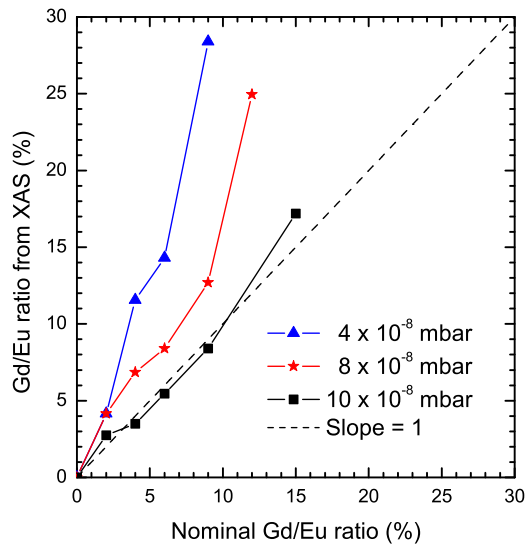}
\caption{\label{Fig_5} (Color online) The actual Gd/Eu ratio in
Gd--doped EuO films determined from the Eu and Gd $M_{4,5}$ XAS
spectra, versus the nominal ratio given by the relative flux
rates. Various oxygen pressures used are indicated and the Eu flux
rate was 8.1--8.2~\AA/min in all cases. The dash line with
slope~=~1 serves as guide to the eyes.}
\end{figure}

We will now address the important issue concerning the
relationship between the actual Gd/Eu ratio in the films as
determined from the XAS experiments and the nominal Gd/Eu flux
rate ratio during the preparation, as measured by the
quartz--crystal monitor. We have discovered the actual Gd/Eu ratio
can deviate strongly from the nominal one depending on the oxygen
pressure used and the flux of Gd deposited. Figure~\ref{Fig_5}
displays an overview of our extensive investigation. One can
clearly see that the deviation is largest for films grown under
low oxygen pressures ($4\times10^{-8}$ mbar) or films having the
highest Gd concentrations. Only for films with not too high Gd
concentrations and prepared with the oxygen pressure tuned close
to the critical value ($10\times10^{-8}$ mbar), one can find that
the actual and nominal Gd/Eu ratios match. These observations can
be well related to the fact that the growth rate is limited by the
oxygen pressure. While the Eu can be readily re--evaporated into
the vacuum from a substrate at a temperature of 400$^{\circ}$C,
the Gd can not since it has a much lower vapor pressure. When the
sum of the Gd and Eu fluxes exceeds that of the oxygen, it will be
the Eu which has to accommodate for the excess, and thus alter the
Gd/Eu ratio.

The oxygen pressure dependence of the actual Gd/Eu ratio can also
be taken as yet another confirmation for the occurrence of the
Eu--distillation process.\cite{sutarto09a} This distillation
process is crucial to avoid the accidental formation of oxygen
deficiencies during growth. The existence of a critical oxygen
pressure and its value can also be directly deduced from the fact
that there are data points which lie on the slope = 1 line in
Fig.~\ref{Fig_5}. Moreover, these data points taken at the
critical oxygen pressure demonstrate that the XAS method is very
consistent with the flux--rate method for the determination of the
Gd/Eu ratio. This in turn adds to the credibility of the XAS as a
reliable quantitative method to determine the actual Gd
concentration in also other growth conditions.

\subsection{Magnetic properties}
\begin{figure}[t]
\includegraphics*[scale = 0.3] {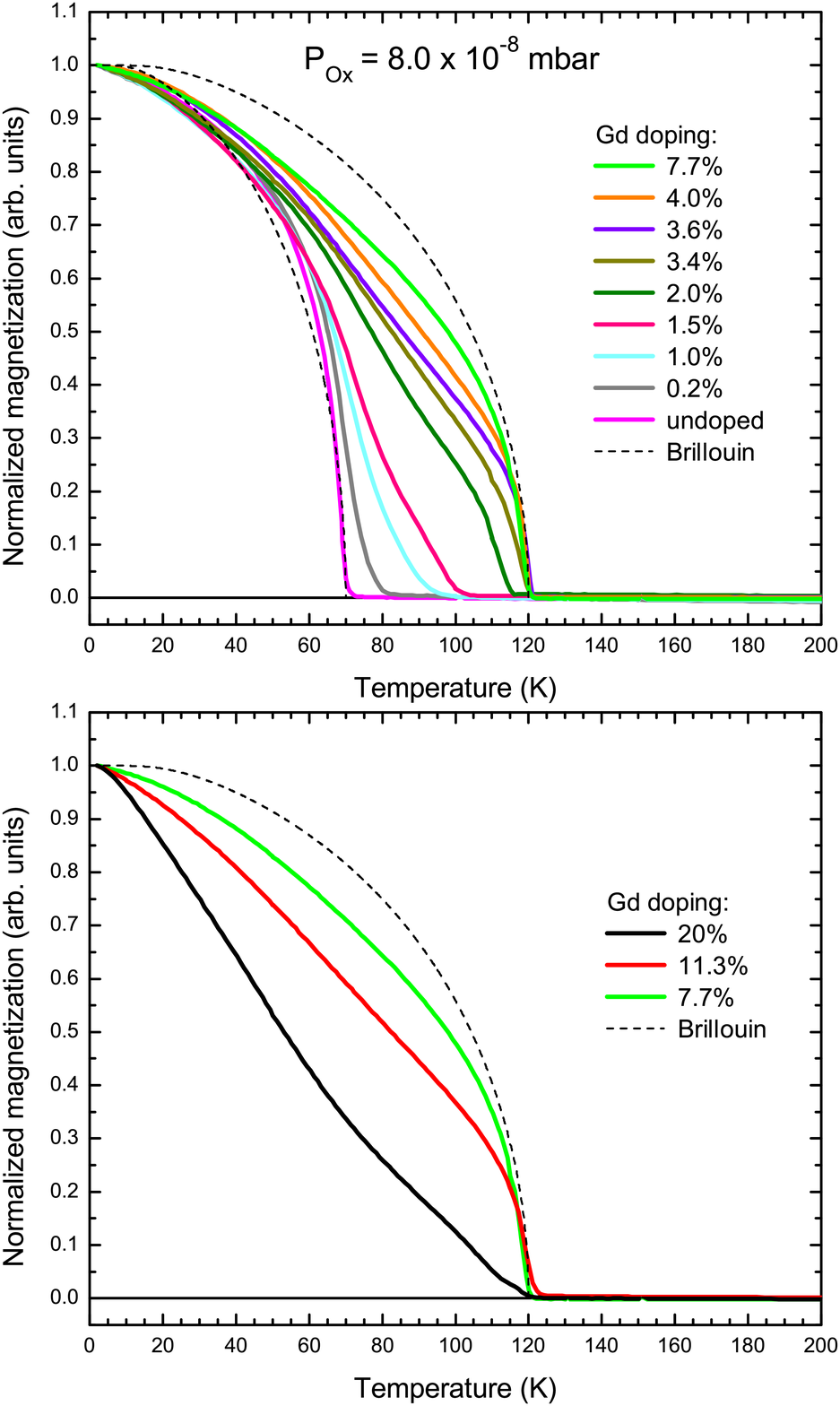}
\caption{\label{Fig_6} (Color online) Temperature dependence of
the normalized magnetization of epitaxial Gd--doped EuO films on
YSZ (001) for various Gd concentrations: panel (a) from undoped to
7.7\% and panel (b) from 7.7\% to 20\%. The applied magnetic field
was 10 G.}
\end{figure}

After having shown that Gd--doped EuO films can be grown with
excellent crystallinity and control of composition, we now focus
our attention on their magnetic properties. Figure~\ref{Fig_6}
depicts the normalized magnetization of a series of Gd--doped EuO
films under an applied magnetic field of 10 G. Starting with the
undoped film, we observe that it has a $T_C$ of 70 K, identical to
that of bulk EuO. The shape of the magnetization curve also
follows the standard Brillouin function quite well, typical for
the undoped bulk EuO.\cite{mauger86a} Upon Gd doping, the $T_C$
increases but the shape of the magnetization curve deviates
strongly from the Brillouin function. The $T_C$ reaches a maximum
of 125 K at 6.5\% doping. For a Gd concentration of 7.7\%, the
magnetization curve comes again closer to the Brillouin function.
The $T_C$ is slightly lower, i.e., 121 K. Further doping with Gd
up to 20\% results in a complete departure of the magnetization
curves from the Brillouin function, as displayed in
Fig.~\ref{Fig_6}(b), but $T_C$ remains $\approx$ 120 K.

\begin{figure}[t]
\includegraphics*[scale = 0.3] {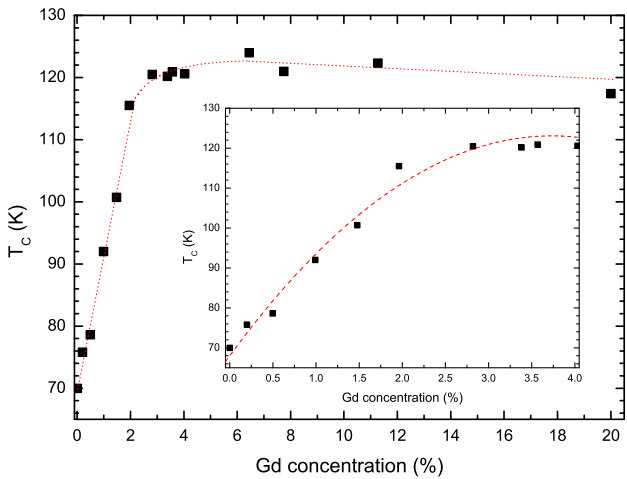}
\caption{\label{Fig_7} (Color online) The Curie temperature of
Gd--doped EuO films on YSZ (001) versus the Gd concentration. The
inset is a close-up for the low Gd concentration range. Dashed
lines serve as guide to the eyes.}
\end{figure}

The $T_C$ dependence on the Gd concentration is presented in more
detail in Fig.~\ref{Fig_7}. We note that the Gd concentration is
the actual concentration in the film as determined by the XAS
method, and not the nominal one based on Gd/Eu flux--rate ratios.
An important result which can be read from Fig.~\ref{Fig_7} is
that only a tiny amount of Gd concentration is needed to enhance
directly the $T_C$. In this respect we cannot confirm the claim
made in the past that the $T_C$ starts only to increase if the Gd
concentration exceeds a threshold value of about
1.2\%--1.5\%.\cite{samokhvalov72a,samokhvalov74a,mauger80a} It is
not clear why the experimental findings are so different. We can
only speculate that in those older studies perhaps not all of the
inserted Gd were substitutional to the Eu and that part of the Gd
could be in the form of, for example, Gd$_2$O$_3$, and thus not
contributing as dopants. In our case we have made the films under
Eu--distillation conditions, i.e., shortage of oxygen, so that it
is very unlikely that Gd$_2$O$_3$ can be formed. In fact, some
other studies were also not able to detect the existence of such a
threshold value.\cite{schoenes74a,matsumoto04a}

Another important result of Fig.~\ref{Fig_7} is that the maximum
value for $T_C$ with Gd doping is 125 K. We cannot reproduce the
higher values (e.g., 130--170 K) reported in the
literature.\cite{shafer68a,ahn68a,samokhvalov72a,samokhvalov74a,mauger78a,meier79a,godart80a,mauger80a,ott06a}
It was inferred that stoichiometry problems could add to an extra
increase in $T_C$.\cite{tsuda91a,schoenes74a} In this respect it
is interesting to note that indeed oxygen deficiencies in Gd--free
EuO samples already lead to enhancements of $T_C$ up to 140--150
K.\cite{massenet74a,samokhvalov78a,matsumoto04a} In our case we
can exclude the presence of oxygen vacancies: the samples all were
grown under Eu--distillation conditions, and under these
conditions we have observed for the pure EuO films that they have
the bulk $T_C$ value of 69 K and that they remain semiconducting
and are, in fact, extremely insulating down to the lowest
temperatures.\cite{sutarto09a}

\begin{figure}[t]
\includegraphics*[scale = 0.3] {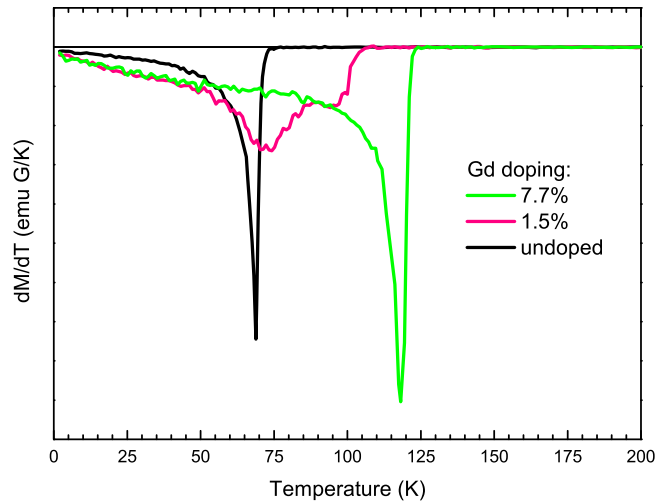}
\caption{\label{Fig_8} (Color online) Temperature derivative of
the magnetization as a function of temperature for the undoped
EuO, the 1.5\% Gd--doped EuO, and the 7.7\% Gd--doped EuO films on
YSZ (001).}
\end{figure}

Our experimental finding of a smooth enhancement of $T_C$ as a
function of Gd concentration is in qualitative agreement with
recent mean--field theoretical models.\cite{ingle08a,arnold08a}
Nevertheless, the magnetization curves deviate strongly from the
Brillouin function upon doping, with the 7.7\% composition showing
the smallest deviation. Such deviations could indicate the
presence of phase separation. To investigate this phenomenon in
more detail, we present in Fig.~\ref{Fig_8} the temperature
derivative of the magnetization curves for the undoped EuO film
and for the 1.5\% and 7.7\% Gd--doped EuO films. One can clearly
see sharp features at 69 K for the undoped film and at about 120 K
for the 7.7\% composition, indicating their relative homogeneity
and corresponding $T_C$'s. For the 1.5\% sample, on the other
hand, one can clearly distinguish two features. Not only there is
a structure at approximately 100 K, marking the $T_C$ of this
film, but there is also a peak at roughly 73 K, which is close to
the $T_C$ of the undoped material.

We now can draw the following picture. Substituting Eu by Gd
results in doping the material with extra electrons, and these
electrons mediate via the double exchange mechanism an enhanced
ferromagnetic coupling between neighboring $4f^7$
ions.\cite{mauger86a,tsuda91a,ingle08a,arnold08a} Nevertheless,
the small impurity potential present at each Gd site binds the
extra electron so that a bound magnetic
polaron\cite{mauger86a,tsuda91a} is formed which becomes
practically ferromagnetic at about 125 K. The diameter of these
bound polarons could be of order three nearest--neighbors
distances. For low doping, they are separated by regions without
Gd doping. In going from high to temperatures lower than 125 K,
each polaron starts to polarize the surrounding undoped regions.
Lowering the temperature further, there will be a temperature at
which a collective ferromagnetic long-range order is created. This
is the $T_C$ of the sample. In the low--doping limit,
nevertheless, a sizable volume fraction of the material does not
feel sufficiently strong the polarizing effect of the paramagnetic
clusters, with the result that this fraction starts to order
magnetically only when the temperature is close to the $T_C$ of
the undoped material. Apparently, this is the case for the 1.5\%
film: the temperature derivative of the magnetization reveals not
only the $T_C$ of 100 K but also another characteristic
temperature at 73 K, see Fig.~\ref{Fig_8}. For higher doping
levels more and more of the volume fraction belongs to the
paramagnetic cluster part and/or gets easily polarized by the
clusters, with the consequence that the $T_C$ increases further
steadily. At 7.7\% concentration one apparently has reached the
situation in which the material is magnetically rather homogenous
as suggested by the fact that the magnetization curve does not
deviate too much from the Brillouin function. For even higher
doping levels the magnetization starts to decrease, the cause of
which is not clear at the moment.

\begin{figure}[t]
\includegraphics*[scale = 0.3] {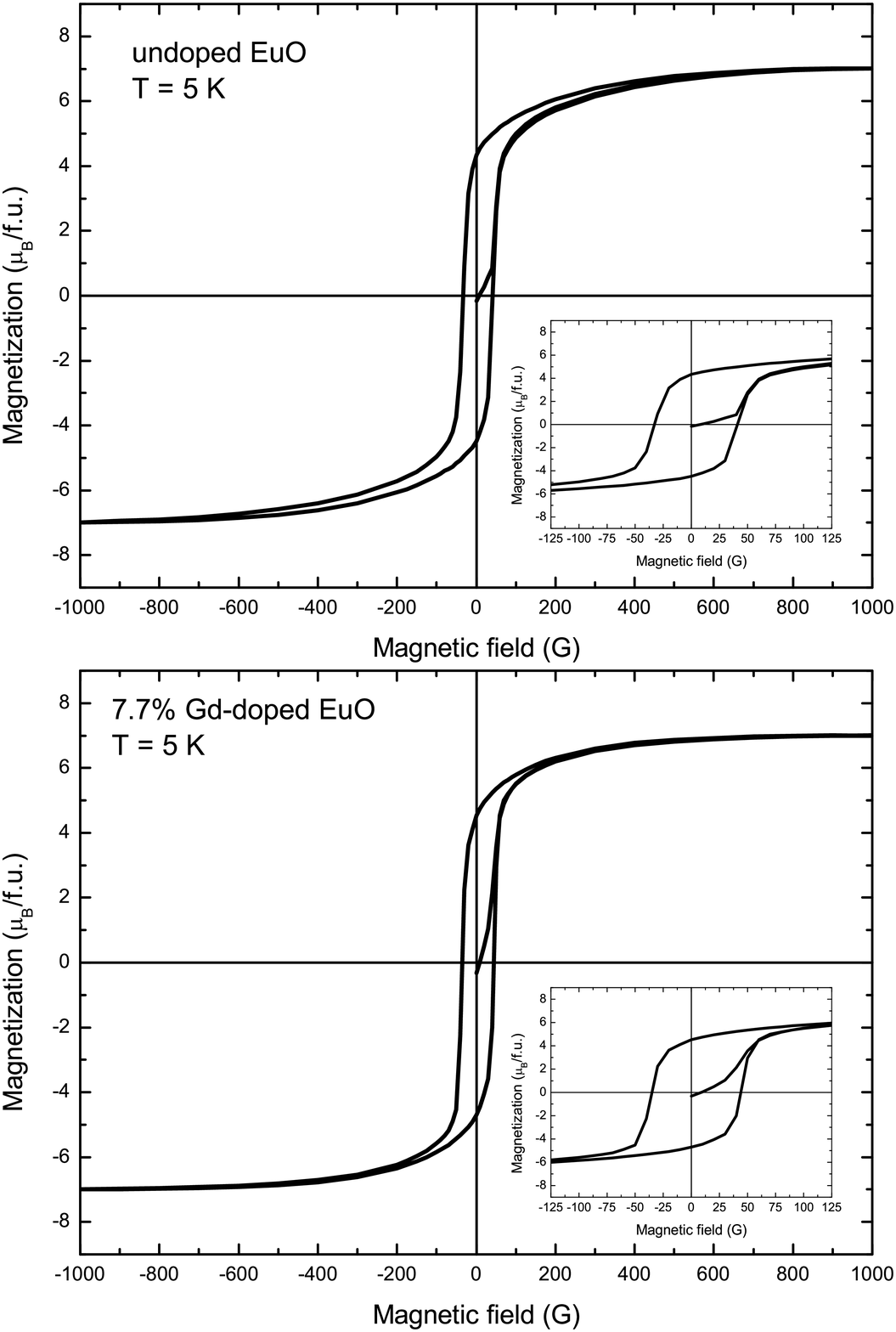}
\caption{\label{Fig_9} The magnetization as a function of the
applied magnetic field of (a) the undoped EuO and (b) the 7.7\%
Gd--doped EuO films on YSZ (001) at 5 K. The insets are
enlargement of the corresponding curves at low applied magnetic
field.}
\end{figure}

In order to elucidate further the magnetic properties, we have
analyzed the magnetization data just below $T_C$ in terms of a
power law, $M(T)$~$\sim$~($T_C$~-~$T$)$^{\beta}$, where $\beta$
denotes the critical magnetic exponent with the typical value of
0.36 for the three--dimensional (3D) Heisenberg model. The small
magnetic contribution from the substrate has been subtracted. We
find $\beta$ of 0.35~$\pm$~0.01 for the undoped EuO film, which is
similar to the bulk EuO\cite{menyuk71a} as a well-known Heisenberg
ferromagnet. Also for the 7.7\% Gd--doped EuO film we extract a
similar $\beta$ value, 0.37~$\pm$~0.01. This means that Gd--doped
EuO behaves also as a Heisenberg ferromagnet. To complete the
investigation of the magnetic properties, we also have measured
the magnetic field dependence of the magnetization for the undoped
EuO and the 7.7\% Gd--doped EuO films at 5 K. The results are
shown in Fig.~\ref{Fig_9}. One can clearly observe that the
samples have very similar hysteresis curves. The remanence
magnetization is about 4.5~$\mu_B$ and the coercive field is about
40~G. Moreover, they also showed the saturation magnetization of
7~$\mu_B$ per f.u. as expected for a $4f^7$ system.

\section{Conclusion}
We have succeeded in preparing high-quality Gd--doped EuO samples.
The films on YSZ (001) are single--crystalline, and a
layer--by--layer epitaxial growth has been observed. Thanks to the
use of the Eu--distillation process during the growth, we were
able to obtain films free from Eu$^{3+}$ species, oxygen
vacancies, and Gd$_2$O$_3$ contaminants. We have provided a
reliable determination of the actual Gd concentration in the films
by applying soft x--ray absorption spectroscopy at the Gd and Eu
$M_{4,5}$ edges. We found that the Curie temperatures increases
steadily as a function of Gd concentration reaching a maximum of
125 K. A threshold behavior was not observed for concentrations as
low as 0.2\%. Both the undoped and the 7.7\% Gd--doped EuO films
reveal magnetic properties typical for a 3D, S~=~7/2, Heisenberg
ferromagnet. For intermediate Gd concentrations we find
indications for phase separation to occur.

\section{Acknowledgments}
We would like to thank Lucie Hamdan and Susanne Heijligen for
their skillful technical assistance. R. S. wishes to thank S.
Standop for his support in the film preparation. We also thank D.
Khomskii for stimulating discussions. We acknowledge the NSRRC
staff for providing us with beam time. The research in Cologne is
supported by the Deutsche Forschungsgemeinschaft through SFB 608.

\end{document}